\documentclass[Physsubmission, Phys]{SciPost}
\hypersetup{
    colorlinks,
    linkcolor={red!50!black},
    citecolor={blue!50!black},
    urlcolor={blue!80!black}
}

\usepackage{subfig}

\usepackage[bitstream-charter]{mathdesign}
\DeclareSymbolFont{usualmathcal}{OMS}{cmsy}{m}{n}
\DeclareSymbolFontAlphabet{\mathcal}{usualmathcal}

\begin{document}

\begin{center}{\Large \textbf{
Can quantized fragmentation replace \\ colour reconnection models ?
}}\end{center}

% TODO: write the author list here. Use initials + surname format.
% Separate subsequent authors by a comma, omit comma at the end of the list.
% Mark the corresponding author with a superscript *.
\begin{center}
\v{S}\'{a}rka Todorova-Nov\'{a} \textsuperscript{$\star$} 
\end{center}

\begin{center}
{\bf } IPNP, Charles University, Prague

* sarka.todorova@cern.ch
\end{center}

\begin{center}
\today
\end{center}

% For convenience during refereeing (optional),
% you can turn on line numbers by uncommenting the next line:
%\linenumbers

\definecolor{palegray}{gray}{0.95}
\begin{center}
\colorbox{palegray}{
  \begin{tabular}{rr}
  \begin{minipage}{0.1\textwidth}
    \includegraphics[width=23mm]{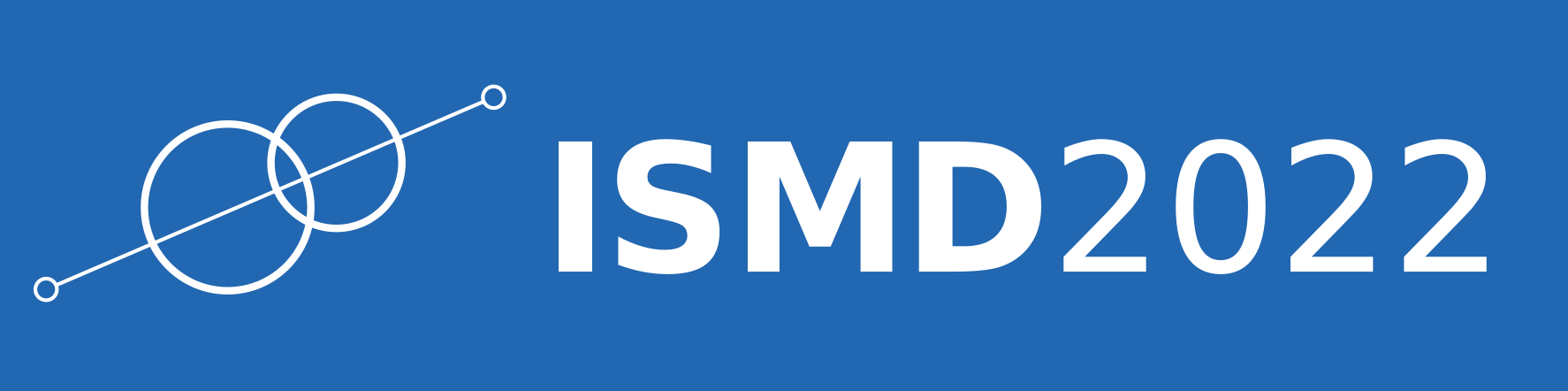}
  \end{minipage}
  &
  \begin{minipage}{0.8\textwidth}
    \begin{center}
    {\it 51st International Symposium on Multiparticle Dynamics (ISMD2022)}\\ 
    {\it Pitlochry, Scottish Highlands, 1-5 August 2022} \\
    \doi{10.21468/SciPostPhysProc.?}\\
    \end{center}
  \end{minipage}
\end{tabular}
}
\end{center}

\section*{Abstract}
{\bf
% TODO: write your abstract here.
A brief overview of properties of the model of quantized fragmentation of a helical QCD string
is followed by the discussion of colour reconnection models and of the ways to reproduce the evolution of the average transverse momentum in hadronic events,
as function of charged particle multiplicity. Reference to recent experimental data is included.
}

%\vspace{10pt}
%\noindent\rule{\textwidth}{1pt}
%\tableofcontents\thispagestyle{fancy}
%\noindent\rule{\textwidth}{1pt}
\vspace{10pt}

\section{Introduction}
\label{sec:intro}
  The notion that correlations between hadrons are an important source of information about hadron formation is widely accepted but not sufficiently explored.  Recently,  an alternative approach to the modelling of the confinement, replacing 1-dimensional string of the Lund fragmentation model by a 3-dimensional helical string, brought some insight into rules governing the hadronization. There are several essential properties the 3-dimensional string brings into consideration: the intrinsic transverse momentum of hadron being defined by the transverse shape of the string, the correlation between intrinsic transverse and longitudinal momentum components are defined.  Of special interest is the possibility to study, for the first time, the causal relations between string breakups which define direct hadrons. The requirement of a cross-talk between endpoint vertices reveals a quantization scheme in which different hadron species correspond to a helical string breaking in regular intervals of helix phase $\Delta\Phi\sim$2.8 rad \cite{helix2,baryons}. The mass spectrum of light hadrons is described, with precision of 1-3\%, with help of only two parameters: $\Delta\Phi$ and the energy scale $\kappa R$, where  $\kappa$ stands for string tension and $R$ for the radius of helix.  Since the quantization proceeds in the transverse mass $m_t = \sqrt{m^2 + p_t^2}$ rather than mass $m$ alone, the model predicts non-trivial correlations between colour-adjacent hadrons. A number of these predictions concerning correlations between colour-adjacent direct hadrons has been validated by recent ATLAS measurements \cite{confnote}, see also \cite{ismd22_chains}. For the purpose of these validations, new observables sensitive to the dynamics of hadron production have been introduced which are of interest for the discussion of colour reconnection.

\section{Observable sensitive to colour flow}
\begin{figure}[h]
\centering
\includegraphics[width=0.48\textwidth]{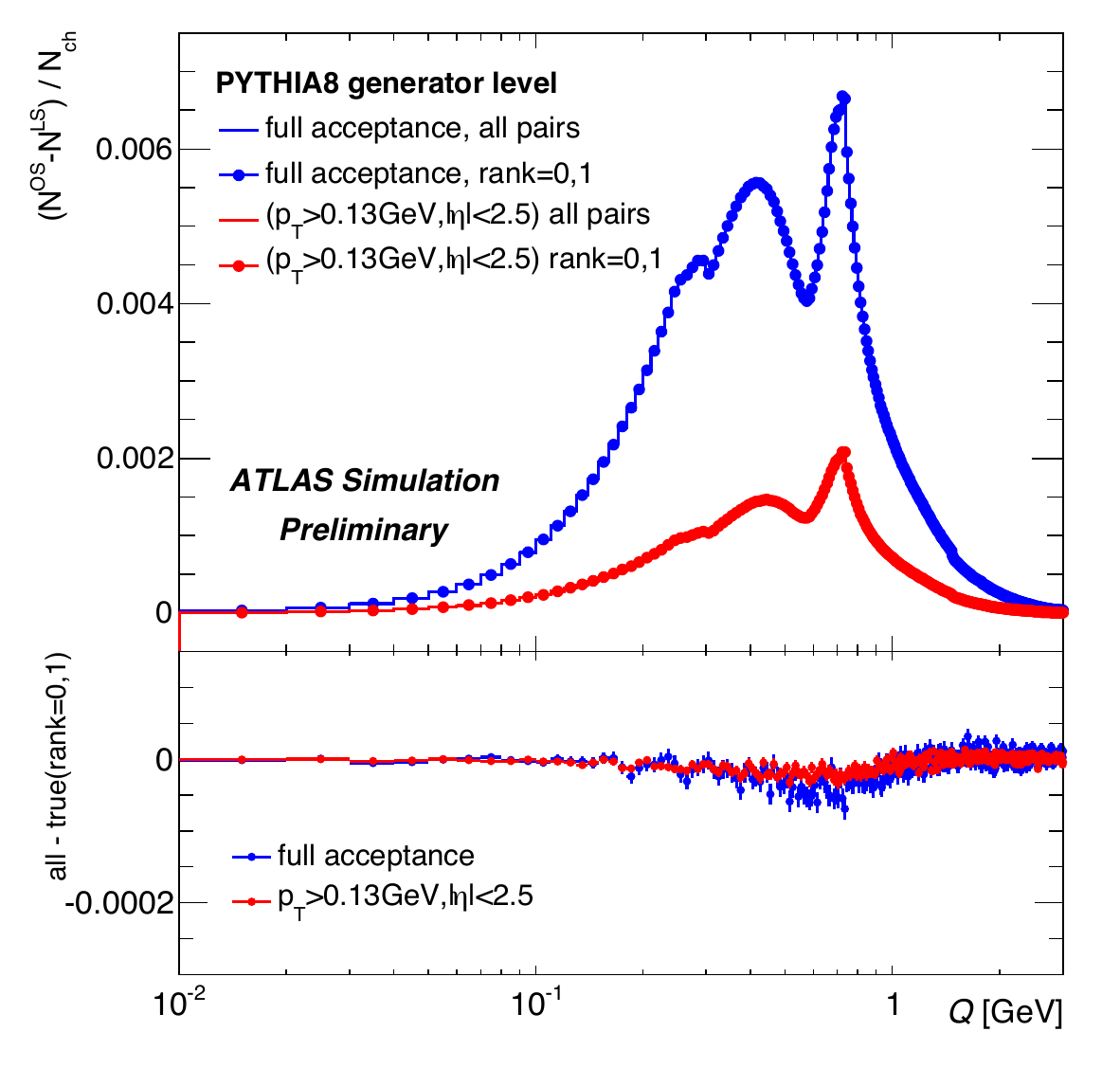}
\includegraphics[width=0.48\textwidth]{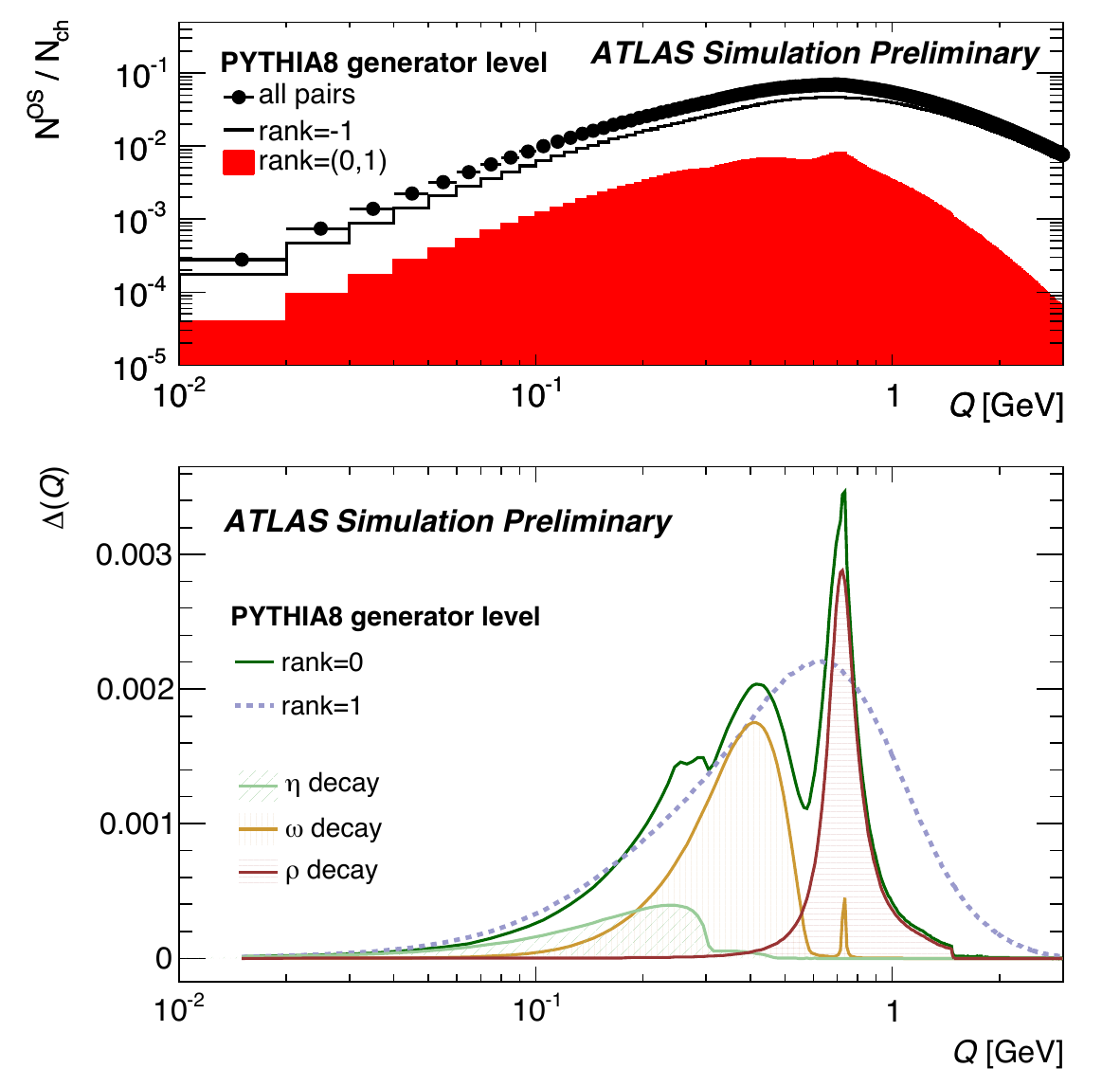}
\caption{ \textsc{PYTHIA8} generator-level study of properties of observable $\Delta(Q)$. Rank (difference) refers to the ordering of hadrons along the colour-flow (decay products inherit rank from their direct parent, rank difference -1 is attributed to hadrons coming from different colour singlet sources.  Left:  There is a close correspondence between the shape of $\Delta(Q)$ and the signature of colour-adjacent hadrons, obtained from the event history record. This feature is independent from acceptance cuts applied on charged particles. Top right: Plot illustrating the size of the signal of colour-adjacent  
pairs as compared to the inclusive OS spectrum, dominated by combination of hadrons from different strings. Bottom right:  Plot showing the principal components of the signature of colour-adjacent hadron pairs. The three-peak structure originates in decays of  $\eta, \omega$ and $\rho$ mesons. Dashed line indicates the distribution of unbound hadron pairs. Source: Ref\cite{confnote}. }
\label{fig:delta_def}
\end{figure}

 Figure \ref{fig:delta_def} shows \textsc{PYTHIA8}\cite{pythia8} generator-level study of properties of observable based on the difference between inclusive spectra of like-sign (LS) and opposite-sign (OS) hadron pairs, normalized to the number of charged particles in the sample: $\Delta(Q)= \frac{1}{N_\mathrm{ch}}  [ N^{OS}(Q) - N^{LS}(Q) ]$.  $\Delta(Q)$ is uniquely sensitive to the momentum difference between colour-adjacent hadrons, which are classified according to pair rank difference r (rank describes the ordering of hadrons according to colour flow): r=0 for decay products of direct hadrons, r=1 for direct hadrons sharing a common string breaking vertex.

\begin{figure}[h]
\centering
\subfloat[]{ \includegraphics[width=0.55\textwidth]{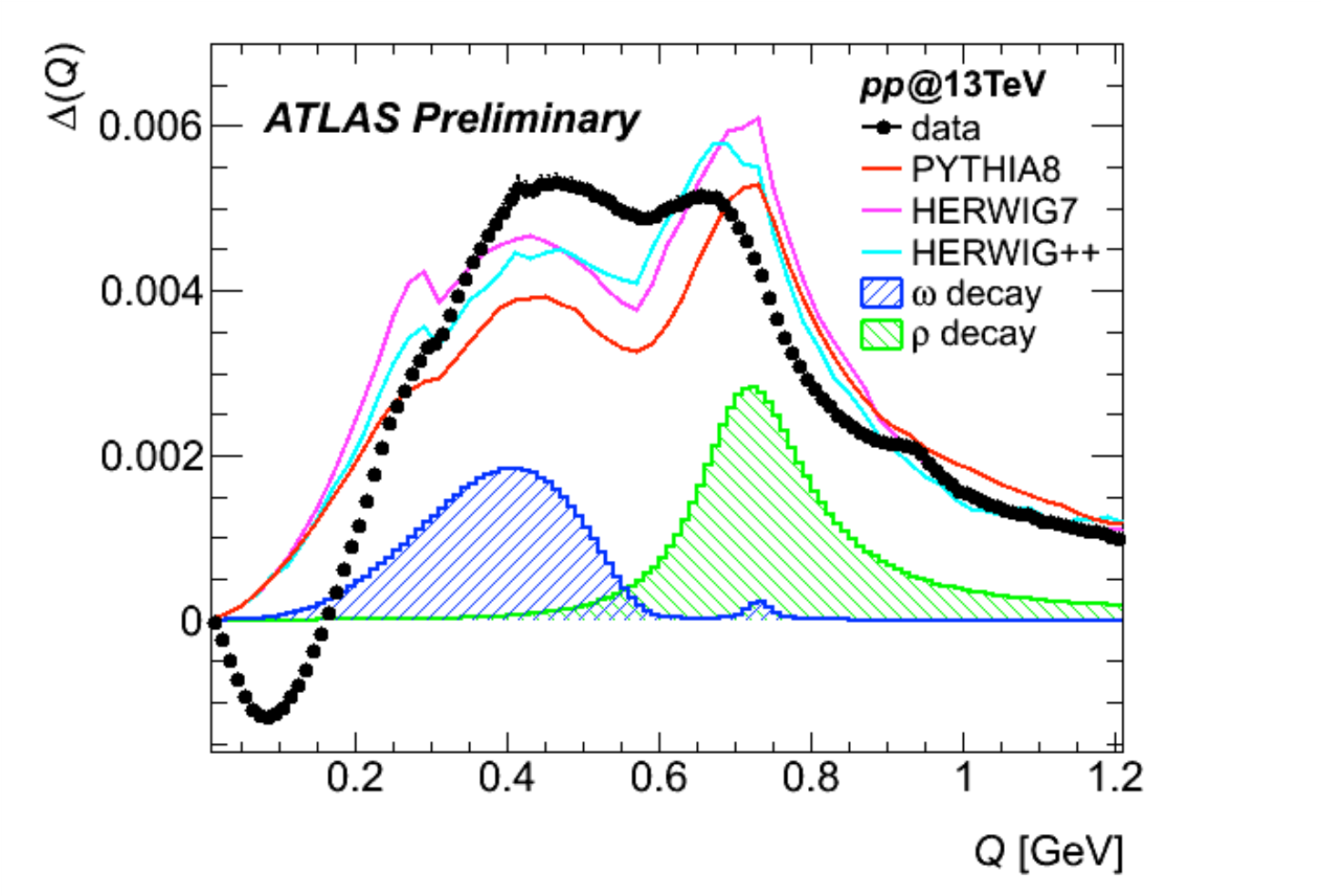} \label{fig:delta_diff_a} }
\subfloat[]{ \includegraphics[width=0.43\textwidth]{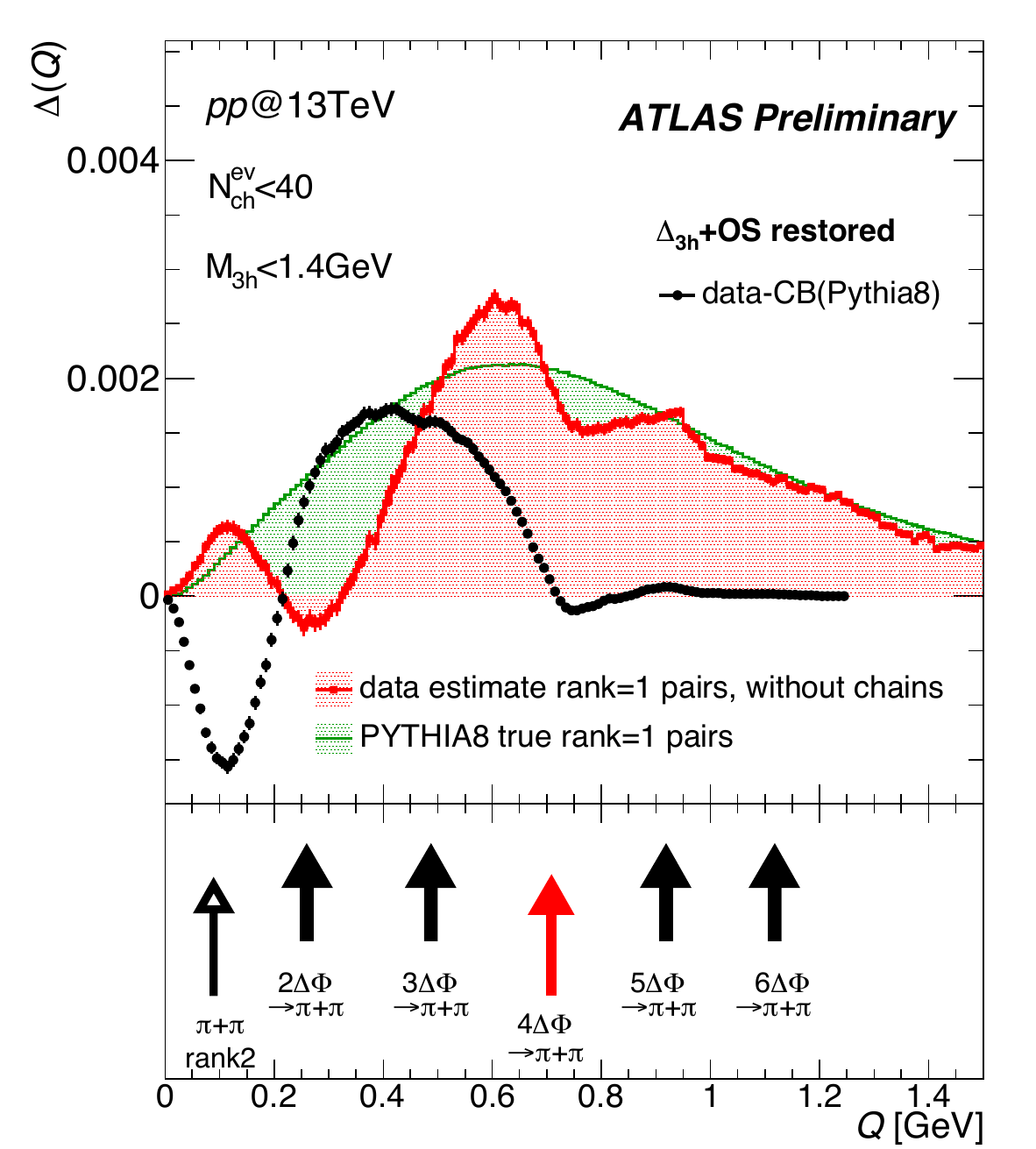} \label{fig:delta_diff_b} }
\caption{ (a) Comparison of $\Delta(Q)$ measured by ATLAS in $pp$@13TeV with conventional hadronization models. (b) Comparison of r=1 distributions between data and \textsc{Pythia8} 
under assumption that decays (r=0) are correctly described by \textsc{Pythia8}.  The signal obtained from triplet chains associated with the anomalous production of LS pairs is subtracted from the measured r=1 spectrum and overlayed. Source: Ref\cite{confnote}.
\label{fig:delta_diff} } 
\end{figure}

 Significant differences appear in the comparison of ATLAS $pp$@13TeV data with predictions of conventional hadronization models, Figure \ref{fig:delta_diff}\subref{fig:delta_diff_a}.
 Anomalous production of LS pairs, visible as a negative part of the $\Delta(Q)$ distribution, is conform to the expectations of the quantized fragmentation  and it is associated with an excess of ordered triplet chains over the \textsc{Pythia8} prediction as described in Ref \cite{ismd22_chains}.  The shape of $\Delta(Q)$ depends on the dynamics of the hadronization as well as on the mass distribution of hadronic sources, none of which seems to be sufficiently well understood. A tentative estimate of differences between r=1 pairs (unbound colour-adjacent) is presented in Figure \ref{fig:delta_diff}\subref{fig:delta_diff_b}. Under assumption that \textsc{Pythia8} describes the decay spectra reasonably well, the generated (r=0) distribution is subtracted from the data and the result is compared with \textsc{Pythia8} r=1 distribution of uncorrelated unbound pairs. Signal from chains associated with anomalous LS pair production
(presumably composed of r=1,2 pairs)  is subtracted as well and overlayed. Data show modulations
which can be tentatively associated with two-pion "decays" of quantized $(1+n)\Delta\Phi$ states.  An interesting
feature of the picture is the absence or even lack of pairs corresponding to $(1+3)\Delta\Phi$ state (indicated by red arrow) : the hypothesis is that the unbound state becomes integrated with the $\rho$ shape. The hypothesis may explain the variation of $\rho$ mass and width measured in $\tau$ decays and in hadroproduction \cite{pdg}.    
 
\begin{figure}[h]
\centering
\subfloat[]{ \includegraphics[width=0.49\textwidth]{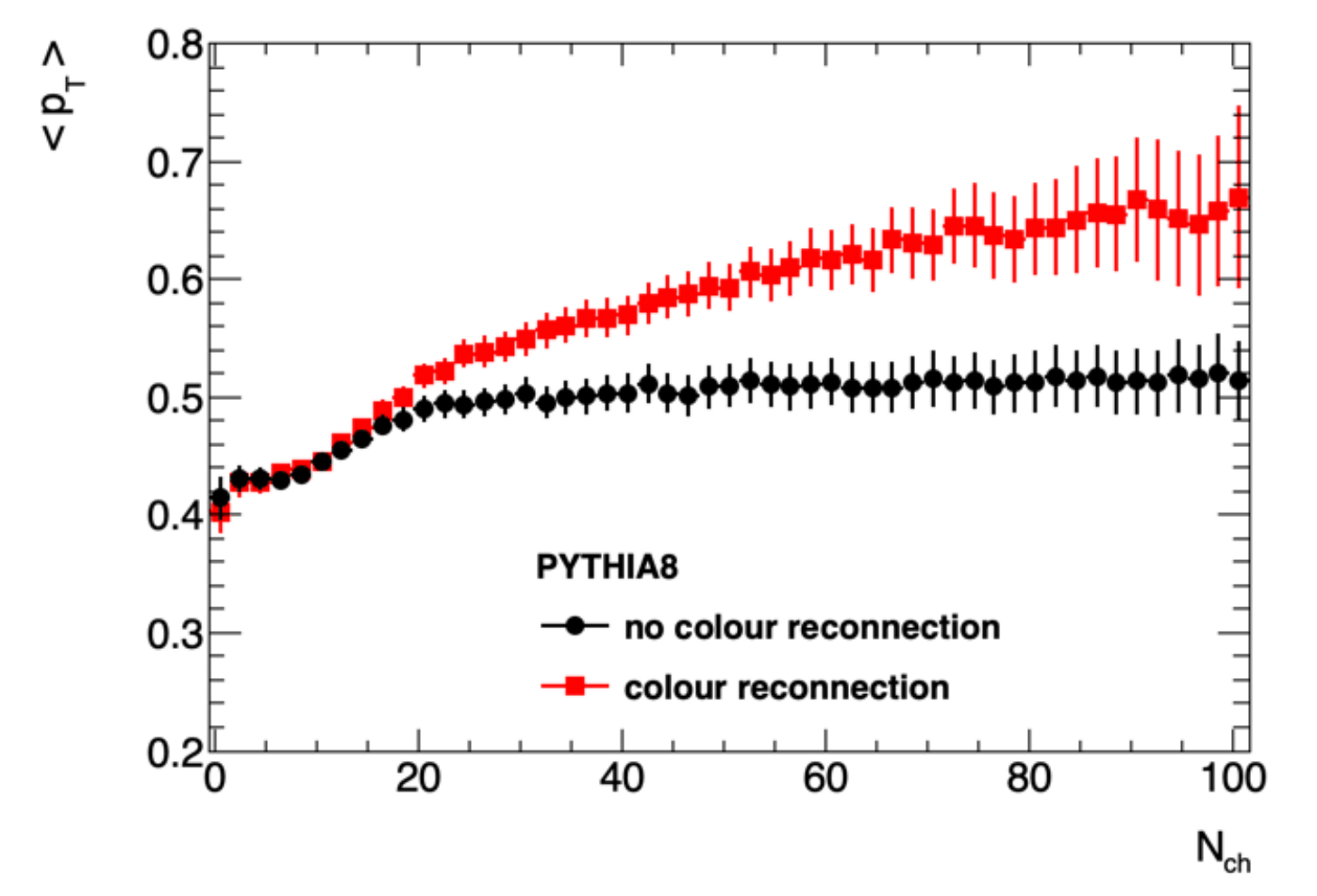} \label{fig:cr_py8_a} }
\subfloat[]{ \includegraphics[width=0.49\textwidth]{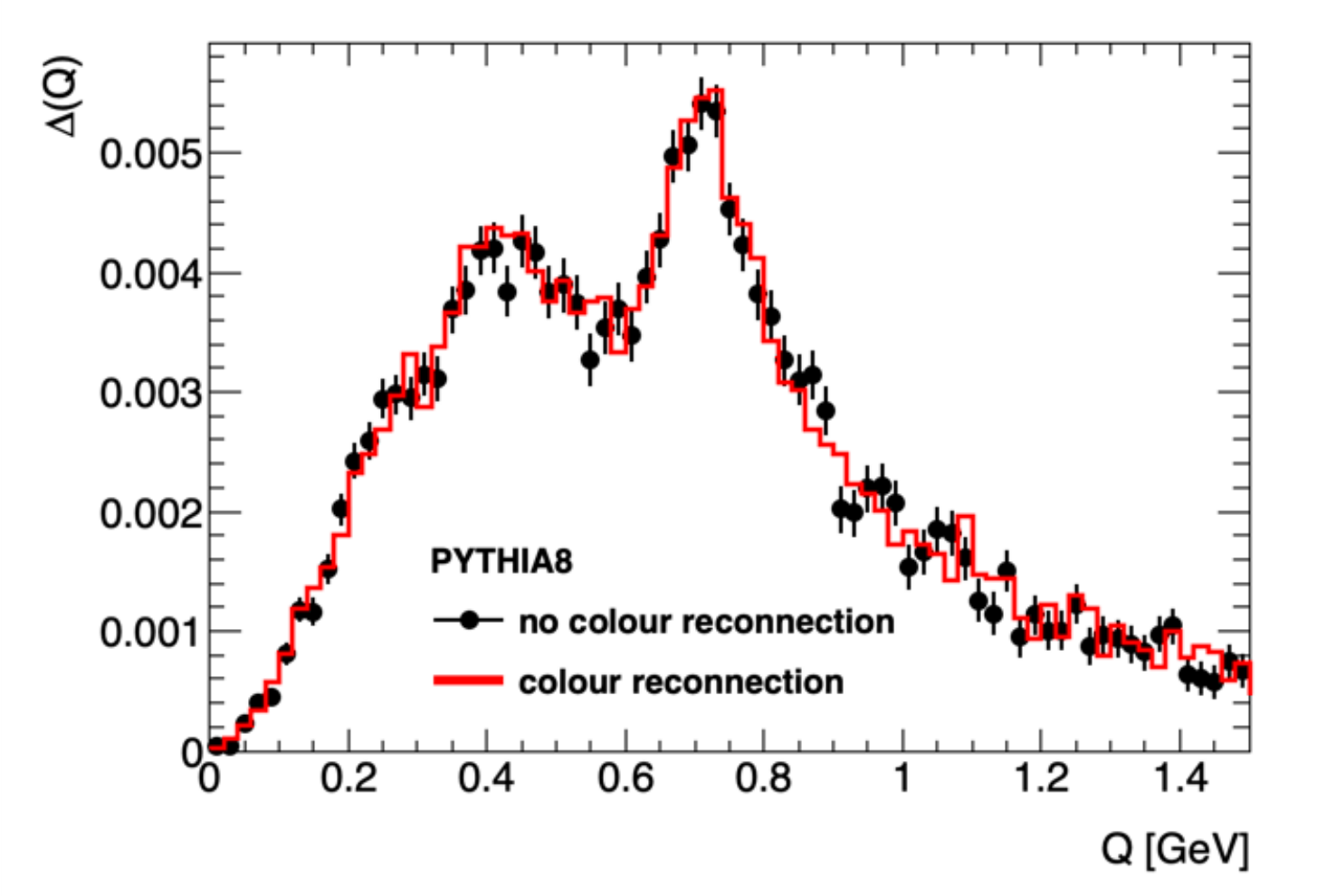} \label{fig:cr_py8_b} }
\caption{ Illustrative \textsc{PYTHIA8} study of the impact of colour reconnection on the average p$_\mathrm{T}$ (in GeV) as function of charged particle multiplicity (a) and $\Delta(Q)$ (b). $pp$ sample generated with 4C tune and colour reconnection switched on and off.
\label{fig:cr_py8} } 
\end{figure}

\section{Colour reconnection}

  Colour reconnection models were developed to study systematic uncertainties related to possible interaction between overlapping hadronic systems in the measurement of WW production at LEP2.
Facing the unknown phenomena, and driven by conservative approach, the use of sometimes extreme changes of colour flow between partons was justified and calibrated by experimental measurements of particle flow outside jets. Later on, colour reconnection models became a popular component of LHC generator tunes, mostly for their impact on the evolution of <p$_\mathrm{T}$> with charged particle multiplicity, Figure \ref{fig:cr_py8}\subref{fig:cr_py8_a}.  The criteria used for the change of colour flow 
are however only weakly motivated by QCD, and the reconnection procedure may be counter-productive as it interferes with the study of QCD motivated evolution of parton showers.  Such a concern is only enhanced by the study of $\Delta(Q)$ observable. Sensitive to colour flow and poorly described as it is, the observable seems not at all affected by colour reconnection algorithm, Figure \ref{fig:cr_py8}\subref{fig:cr_py8_b} (a "generic" example of colour reconnection algorithm is used for the illustration of the argument). 

An ad-hoc modification of the colour flow which does not fix an obvious problem in the description of the colour flow seen in Figure \ref{fig:delta_diff}\subref{fig:delta_diff_a} seems indeed apt to increase the confusion and to complicate further QCD studies. It should be acknowledged that colour reconnection models do not modify the hadronization with the aim to introduce correlations between colour-adjacent hadrons observed directly and indirectly in the data. Instead, they modify mass spectrum of initial hadronizing sources, but the range of sources with low masses which is of importance for the description of $\Delta(Q)$ observable seems not to be correctly addressed \cite{confnote}.

\begin{figure}[h]
\centering
\subfloat[]{ \includegraphics[width=0.54\textwidth]{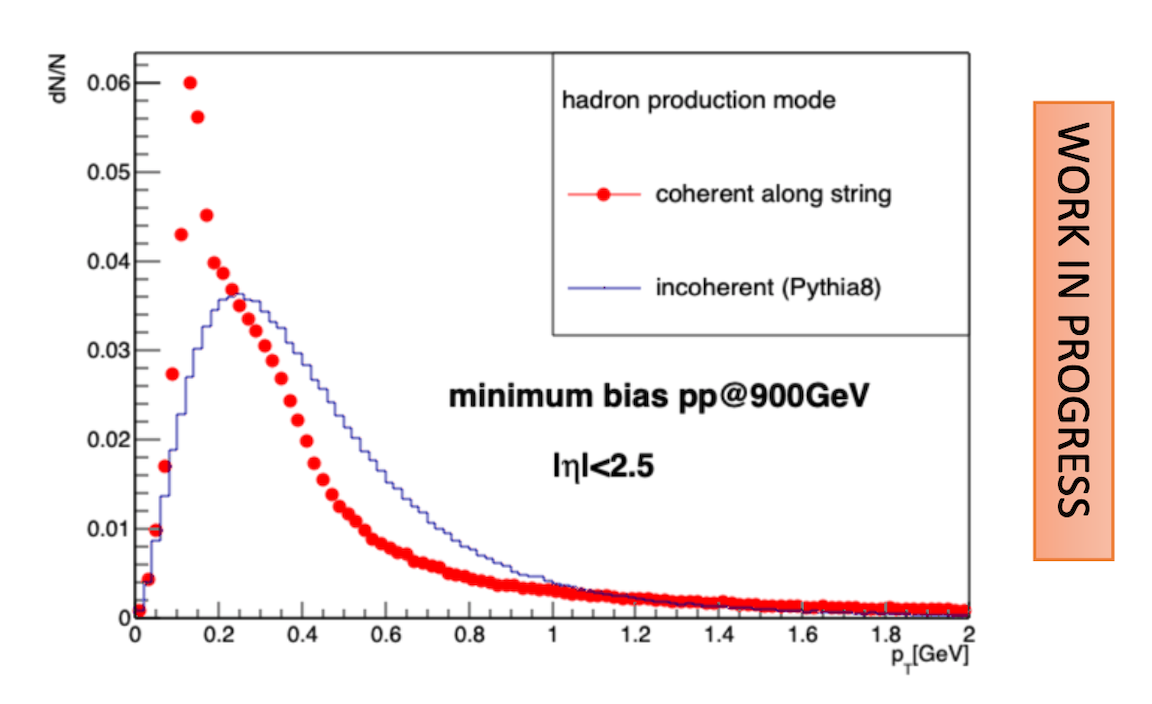} \label{fig:helix_cr_a} }
\subfloat[]{ \includegraphics[width=0.44\textwidth]{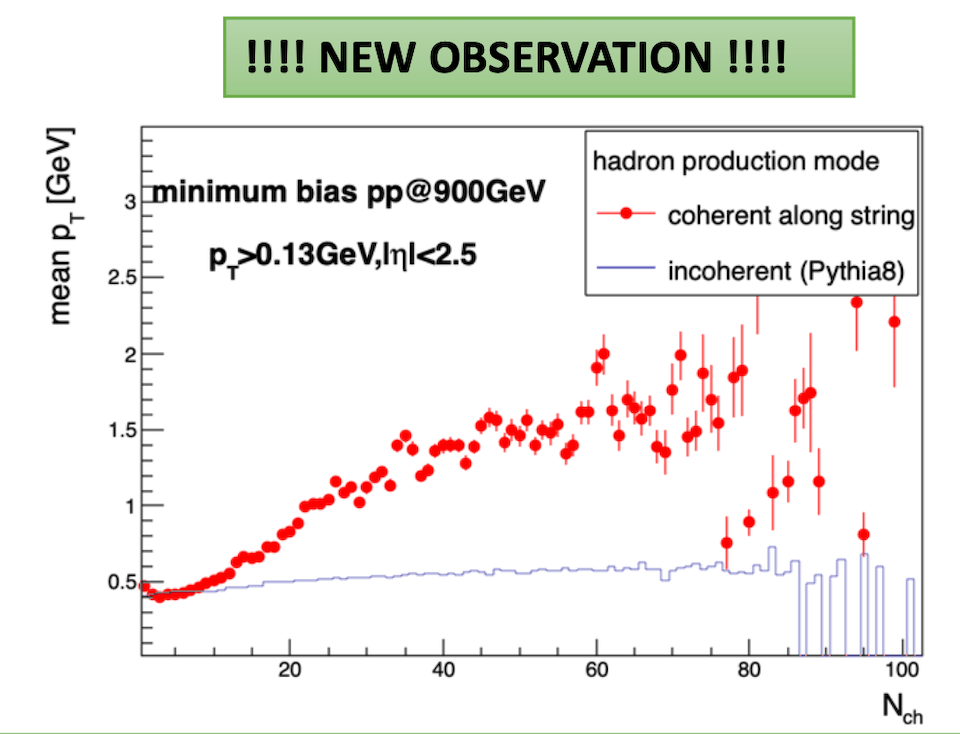} \label{fig:helix_cr_b} }
\caption{Monte-Carlo study of extreme scenario of hadron production by fully coherent fragmentation of homogeneous helical string.  (a)  Lowering of the p$_\mathrm{T}$
cut-off to 50 MeV should allow to measure at least 99\% of final charged particles filtered by $\eta$ acceptance of ATLAS tracker. (b) A strong impact of the quantized fragmentation on the average p$_\mathrm{T}$ is revealed, here plotted as a function of charged particle multiplicity.}
\label{fig:helix_cr}
\end{figure}

\section{Average transverse momentum in quantized fragmentation}
   The model of quantized helical string does not - at the time of writing - constrain the fragmentation function (the evolution of the pitch of helix along the string axis). 
 Figure \ref{fig:helix_cr} shows an extreme scenario of quantized fragmentation of a homogeneous string
 where all hadrons are produced in a fully coherent manner, by induced gluon splitting \cite{baryons}.
 The study indicates the region most sensitive to the signature of quantized fragmentation 
 (~$\sim$50 MeV < p$_\mathrm{T}$ < $\sim$100 MeV) - and it also reveals a strong variation 
  of <p$_\mathrm{T}$> with the particle multiplicity.  This observation shows that colour reconnection
 can be, in principle, replaced by modeling which preserves the "natural" colour flow of the event.
 As in the case of the study of anomalous production of LS pions \cite{ismd22_chains}, it seems
 important to find experimental setup demonstrating the link - if it exists - between correlations
   ($\Delta(Q)$) 
 and the evolution of inclusive particle spectra (<p$_\mathrm{T}$>(N$_\mathrm{ch}$)) in order to
understand the longitudinal properties of the helical string.  From the phenomenological point of view, this can be also viewed as a question of modeling of late stages of the development of parton shower. This returns the model to its origins - the use of helical string has been originally suggested by Lund theorists \cite{lund_helix} as a way to stabilize the end of parton cascade.

%  Partial implementation of the model in MC studies of systematic uncertainties in the analysis of the %data however shows a strong impact of   helical ordering of hadrons on the average transverse %momentum in the minimum bias sample.

\section{Conclusion}
 With help of recently introduced observable which is exceptionally sensitive to the dynamics of hadron production,  it is argued that colour reconnection models may be detrimental to the understanding of various phenomena contributing to the description of hadronic final states: number and mass spectrum of hadron sources, partons showers and hadronization.  It is shown that development of quantized fragmentation model should resolve at least part of discrepancies between data and conventional models which justify the widespread use of colour reconnection models.

%\section*{Acknowledgements}
%Acknowledgements should follow immediately after the conclusion.

% TODO: include author contributions
%\paragraph{Author contributions}
%This is optional. If desired, contributions should be succinctly described in a single short paragraph, using author initials.

% TODO: include funding information
\paragraph{Funding information}
This work was partially supported by the Inter-Excellence/Inter- Transfer grant LTT17018 and the Research Infrastructure project LM2018104 funded by Ministry of Education, Youth and Sports of the Czech Republic, and the Charles University project UNCE/SCI/013.

%\begin{appendix}
%\section{First appendix}
%Add material which is better left outside the main text in a series of Appendices labeled by capital letters.

% TODO:
% Provide your bibliography here. You have two options:

% FIRST OPTION - write your entries here directly, following the example below, including Author(s), Title, Journal Ref. with year in parentheses at the end, followed by the DOI number.
%\begin{thebibliography}{99}
%\bibitem{1931_Bethe_ZP_71} H. A. Bethe, {\it Zur Theorie der Metalle. i. Eigenwerte und Eigenfunktionen der linearen Atomkette}, Zeit. f{\"u}r Phys. {\bf 71}, 205 (1931), \doi{10.1007\%2FBF01341708}.
%\bibitem{arXiv:1108.2700} P. Ginsparg, {\it It was twenty years ago today... }, \url{http://arxiv.org/abs/1108.2700}.
%\end{thebibliography}

% \bibliographystyle{SciPost_bibstyle} % Include this style file here only if you are not using our template
%\bibliography{SciPost_Example_BiBTeX_File.bib}

%\nolinenumbers

\end{document}